\documentclass[a4paper,11pt]{article}
\usepackage{jcappub} 
\usepackage{lineno}
\usepackage{graphicx}
\usepackage{dcolumn}
\usepackage{bm}
\usepackage{mathtools}
\usepackage{url}
\usepackage{graphicx, subfigure}
\usepackage{amsmath,amssymb}
\usepackage{hyperref}
\usepackage[normalem]{ulem}
\usepackage{caption}

\title{Gravitational wave lensing: probing Fuzzy Dark Matter with LISA}

\author[a,b]{Shashwat Singh}
\affiliation[a]{SUPA, University of Glasgow, Glasgow, G12 8QQ, UK}
\affiliation[b]{Observatoire de Paris, Université PSL, 61 avenue de l’observatoire 75014 Paris, France}
\emailAdd{s.singh.3@research.gla.ac.uk}

\author[c,d]{Guilherme Brando}
\affiliation[c]{CBPF- Brazilian Center for Research in Physics, Xavier Sigaud st. 150, zip 22290-180, Rio de Janeiro, RJ, Brazil}
\affiliation[d]{Durham University, Stockton Rd, Durham, DH1 3LE, UK}
\emailAdd{gbrando@cbpf.br}

\author[e]{Stefano Savastano}
\affiliation[e]{Max Planck Institute for Gravitational Physics (Albert Einstein Institute), Am Mühlenberg 1, D-14476 Potsdam-Golm, Germany}
\emailAdd{stefano.savastano@aei.mpg.de}

\author[e]{Miguel Zumalacárregui}
\affiliation[e]{Max Planck Institute for Gravitational Physics (Albert Einstein Institute), Am Mühlenberg 1, D-14476 Potsdam-Golm, Germany}
\emailAdd{miguel.zumalacarregui@aei.mpg.de}

\abstract{Gravitational lensing is a universal phenomenon: it affects both gravitational waves (GWs) and electromagnetic signals travelling through the gravitational field of a massive object. In this work, we explore the prospects of observing lensed GW signals from the mergers of massive black holes, lensed by dark matter halos composed of Fuzzy Dark Matter (FDM), which form dense cores known as solitons. We focus on wave optics phenomena, where frequency-dependent signatures can be observed in the weak lensing regime (i.e. single-image). Our results show that lensing diffraction signatures differ for low-mass halos in FDM, and can reveal the presence of a solitonic core. Furthermore, we demonstrate that FDM and cold dark matter profiles can be distinguished in GW signals from binary massive black hole mergers, which will be observed by the Laser Interferometer Space Antenna (LISA) mission. However, the dense solitonic core does not substantially enhance the detectability of FDM halos at large source-lens offsets, relative to standard cold dark matter. Our analysis confirms FDM halos as a promising signature of dark matter on GW observations.}

\begin{document}
\maketitle

\flushbottom
\section{Introduction}
Gravitational waves (GWs) undergo lensing, similar to electromagnetic (EM) waves when intervened by gravitational fields (e.g., such as those created by massive objects). GWs experience deviation and magnification similar to EM waves. Because of the much longer wavelengths of GWs as compared to EM waves, wave optics (WO) effects such as diffraction and interference become prominent. These WO effects are crucial in studying lens features, lost in the high-frequency geometric optics (GO) regime. The frequency-dependent perturbations in the GWs carry rich information about the density distribution of the lens \cite{Takahashi_2003,tambalo2022lensing}, allowing for accurate lens reconstruction. Moreover, GWs weakly interact with matter and are coherent. Coherence makes GWs sensitive to GW amplitude, which scales as inverse distance compared to EM wave, which falls as inverse-square, making sources at higher redshifts detectable. Taking advantage of the nature of GWs in this regime, it is possible to obtain detailed insights into the properties of gravitational lenses \cite{Savastano_2023, Fairbairn_2023, tambalo2022gravitational, tambalo2022lensing, _al_kan_2023, Lin:2023ccz, Urrutia:2021qak, Urrutia:2023mtk, Caliskan:2022hbu, Caliskan:2023zqm, Choi:2021bkx, GilChoi:2023ahp, Mukherjee:2019wcg, Mukherjee:2019wfw}. 

In this work, we study the effect of GW lensing due to dark matter halo modelled as fuzzy dark matter (FDM) \cite{Hu_2000, schive2014cosmic, Hui_2017} (see Refs.~\cite{Ferreira:2020fam, Hui:2021tkt} for recent reviews), an alternative to cold dark matter (CDM). FDM resolves some of the subgalactic tensions with CDM (discussed in Sec.~\ref{FDM}) \cite{Bullock_2017, Del_Popolo_2017, Bull_2016}, such as the core cusp problem \cite{Dubinski1991496, de_Blok_2009, Walker_2011, WeinbergBullockGoverntro}, overprediction of high luminous galaxies \cite{Boylan_Kolchin_2012} and that of satellite galaxies \cite{Moore_1999, Klypin_1999}, effect of baryonic physics \cite{McGaugh_2011} and dynamics of central black holes~\cite{Boey:2025qbo}.

Although lensing of electromagnetic sources in FDM leads to rich signatures~\cite{Laroche:2022pjm, Powell:2023jns, Amruth:2023xqj, Diego:2024cez}, we restrict ourselves to GWs in the WO regime only, which can be observed in weakly lensed signals, potentially leading to more frequent detections \cite{Gao_2022,Savastano_2023, Caliskan:2023zqm}. 

Focussed on harnessing the WO features, we work with GW signals from the massive black holes (MBHs) binaries which will be detected by Laser Interferometer Space Antenna (LISA). GW signals from such sources are expected to be one of the loudest sources in the LISA band, with signal-to-noise ratios (SNRs) ranging from a few tens to a few hundred \cite{colpi2024lisadefinitionstudyreport}. 

Our results show that FDM halos ({$10^5-10^{5.5}\textrm{M}\odot$}) composed of boson mass, $m_\phi \simeq 10^{-23}$ {eV} deviate significantly in WO features from the CDM halo modelled as Navarro-Frenk-White (NFW). As halo mass or boson mass increases, the WO features converge to the CDM halo. The structure of this work is organized as follows. In Sec.~\ref{theory}, we establish the basic lensing formalism for propagating GWs where we define the relevant terms, quantities, assumptions, and observables that form the basis of our study. With the fundamental formalism of GW lensing in place, we briefly discuss FDM  in Sec.~\ref{FDM}. Here, we focus solely on the features that separate FDM from CDM and how its quantum nature alleviates some of the sub-galactic tension associated with CDM. We further delve into the formation of an FDM halo (Sec.~\ref{form_FDM}) and the astrophysical constraints (Sec.~\ref{const_FDM}). In the further subsections (Sec.~\ref{modelling-FDM}), we detail how we model our FDM halo and discuss its lensing features in Sec.~\ref{wave_optics_FDM}. Finally, in Sec.~\ref{detection}, we focus on the possible detection of these WO features in the LISA band and conclude in Sec.~\ref{conclusion}.

\section{Gravitational Wave Lensing in the Wave-optics Regime}
\label{theory}
\begin{figure}
    \centering
    \includegraphics[height=6cm, width=8cm]{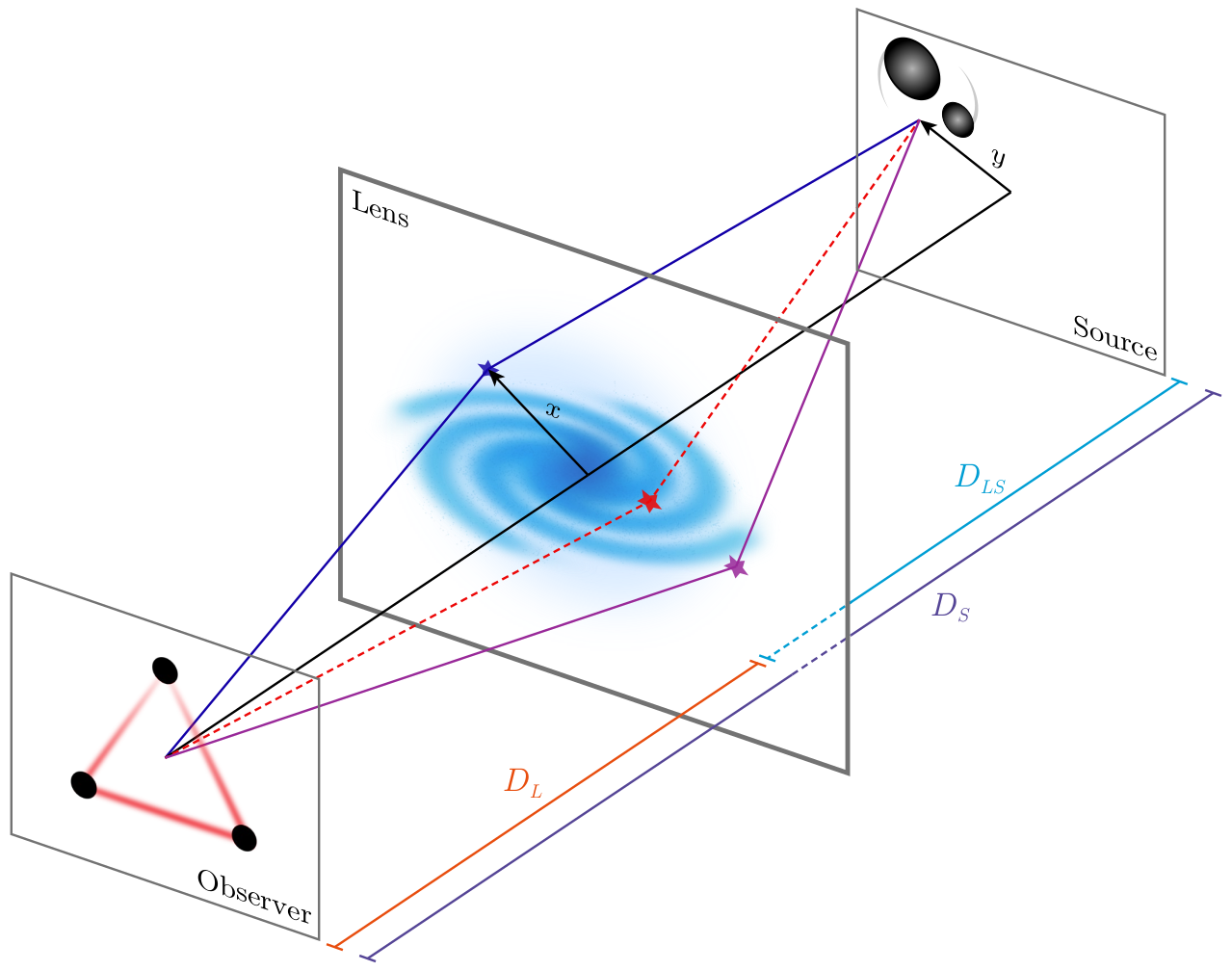}
    \caption{This configuration for gravitational lensing involves a lens positioned at an angular diameter distance $D_L$, which acts to magnify a source located at an angular distance $D_S$. The source is not at the centre of the lens and has an impact parameter $\boldsymbol{y}$. When considering the geometric optics limit, the lens produces multiple images, shown as ``stars'', whose positions are determined by dimensionless coordinates $\boldsymbol{x}=\boldsymbol{\xi}/\xi_0$, in the lens plane. It is possible to observe all the images of the lensed gravitational waves (GWs) signal. [Image: \cite{tambalo2022gravitational}]} 
    \label{GW-lens-theory}
\end{figure}

We begin with the formalism to study lensed GW signals for a general configuration, as shown in {Fig.~\ref{GW-lens-theory}} (for more details on the framework, see Ref.~\cite{villarrubiarojo2024glownovelmethodswaveoptics}). 

The GW source and lens are at angular diameter distance $D_S$ and $D_L$ from the observer, respectively. $D_{LS}$ is the angular diameter distance between the lens and the source. The source centre is not perfectly aligned with the lens centre and has an angular deviation $\boldsymbol{\eta}$, the impact parameter. Under the geometric optics limit, the lens produces multiple images (marked with ``stars'') at dimensionless coordinates $\boldsymbol{x}=\boldsymbol{\xi}_I/\xi_0$ in the lens plane. We develop our formalism under the thin lens approximation and project the density along the line of sight onto the lens plane. The cosmological quantities are defined as follows: scale factor $a = (1+z_L)^{-1}$ and effective distance $d_\mathrm{eff} \equiv aD_LD_{LS}{D_S}^{-1}$. We fix the $\Omega_\Lambda$ and $\Omega_m$ at 0.7 and 0.3, respectively, with the Hubble parameter to be $H_0=70\:\textrm{km/s/Mpc}$ (close to Planck values \cite{2018Planck}) and work in units with $c=1$. 

The quantity of interest is the amplification factor $F$, which is the ratio of lensed to unlensed GW strain (in frequency domain) can be represented as in Eq.~\ref{ampl-fac},
\begin{equation}
\label{ampl-fac}
    F(f) = \frac{\tilde{h}_L(f)}{\tilde{h}_0(f)}
\end{equation}
where $\tilde{h}_L$ and $\tilde{h}_0$ are the lensed and unlensed GW strain in the frequency domain.

The GW strain can be expressed as $h_{\mu\nu} = \Phi e_{\mu\nu}$ where $\Phi$ is the scalar part of the wave and  $e_{\mu\nu}$ is the polarization tensor \cite{takahashi2004wave}. In a case where the wavelength of a wave is much smaller than the typical radius of the curvature of the background, polarization tensor $e_{\mu\nu}$ is parallel-transported along the null geodesic \cite{tambalo2022gravitational}. Thus, we can restrict only to the scalar part ($\Phi$) of the wave. The amplification factor can now be expressed in terms of this scalar part as Eq.~\ref{diff-int}
\begin{equation}
\label{diff-int}
    F(f) = \frac{\tilde\Phi_L}{\tilde\Phi_0}= \frac{-if}{d_\mathrm{eff}}\int d^2\boldsymbol{\xi} e^{2\pi i f t_d(\boldsymbol{\xi},\boldsymbol{\eta})},
\end{equation}
where $t_d(\boldsymbol{\xi},\boldsymbol{\eta})$ is the time delay function of the lens given by Eq.~\ref{time-delay}
\begin{equation}
\label{time-delay}
    t_d(\boldsymbol{\xi},\boldsymbol{\eta})=\frac{1}{2d_\mathrm{eff}}\bigg(\boldsymbol{\xi} - \frac{D_L}{D_S}\boldsymbol{\eta}\bigg)^2 - \hat{\psi}(\boldsymbol{\xi}) - \hat{\phi}_m(\boldsymbol{\eta})
\end{equation}
where $\hat{\psi}(\boldsymbol{\xi})$ and $\hat{\phi}_m(\boldsymbol{\eta})$  are the lensing potential and time delay phase, respectively. Lensing potential can now be computed using 
\begin{equation}
\label{lens-potential}
     \nabla^2_{\boldsymbol{\xi}}\hat{\psi}(\boldsymbol{\xi})=8\pi G \Sigma(\boldsymbol{\xi}),
\end{equation}

where $\nabla^2_{\boldsymbol{\xi}}$ is the Laplacian 2D operator, and $G$ is the Newtonian gravitational constant. In Eq.~\ref{lens-potential}, $\Sigma(\boldsymbol{\xi})$ the projected density is computed by projecting the density along the line of sight of the lens onto the lens plane. For a given density profile {$\rho(r)$ (where $r=\sqrt{\boldsymbol{\xi}^2 + \mathcal{Z}^2}$)} of the lens, the projected density $\Sigma(\boldsymbol{\xi})$ can be expressed as 

\begin{equation}
\label{sigma}
    \Sigma(\boldsymbol{\xi}) = \int_{-\infty}^{\infty}d\mathcal{Z} \:\rho(\sqrt{\boldsymbol{\xi}^2+\mathcal{Z}^2}).
\end{equation}

Eq.~\ref{ampl-fac}$-$\ref{sigma} can be computed in terms of dimensionless quantities $\boldsymbol{x}$ and $\boldsymbol{y}$ defined as in Eq.~\ref{x-y-scaled} as 
\begin{equation}
\label{x-y-scaled}
    \begin{split}
        \boldsymbol{x}=\frac{\boldsymbol{\xi}}{\xi_0} \\
        \boldsymbol{y}=\frac{\boldsymbol{\eta}}{\eta_0}
    \end{split}
\end{equation} 
where $\xi_0$ is an angular scale and $\eta_0 \equiv D_S\xi_0/D_L$. Now, the lensing potential $\hat{\psi}(\boldsymbol{\xi})$ and time delay ${t_d(\boldsymbol{\xi},\boldsymbol{\eta})}$ can also be expressed in terms of dimensionless quantities using the scaling as in Eq.~\ref{x-y-scaled} 
\begin{equation}
\label{phi-psi-scaled}
    \begin{split}
        \phi(\boldsymbol{x},\boldsymbol{y}) = \frac{d_\mathrm{eff}}{\xi_0^2}t_d(\boldsymbol{x},\boldsymbol{y}) \\
        \psi(\boldsymbol{x}, \boldsymbol{y}) = \frac{d_\mathrm{eff}}{a\xi_0^2}\hat{\psi}(\boldsymbol{x},\boldsymbol{y}) \\
    \end{split}
\end{equation}
The rescaled lensing potential can be evaluated as
\begin{equation}
    \nabla^2_{\boldsymbol{x}}\hat{\psi}(\boldsymbol{x})=2\kappa(\boldsymbol{x})
\end{equation}
where $\kappa(x)$ is the convergence given by 
\begin{equation}
\label{kappa-scaled}
    \kappa(\boldsymbol{x}) = \frac{\Sigma(\xi_0\boldsymbol{x})}{\Sigma_\mathrm{cr}}
\end{equation}
with \emph{critical density} $\Sigma_\mathrm{cr} \equiv a(4\pi G d_\mathrm{eff})^{-1}$.

We focus on using Green's function method to evaluate the rescaled lensing potential, given as in Eq.~\ref{psi-scaled}
\begin{equation}
\label{psi-scaled}
    \psi(\boldsymbol{x}) = \frac{1}{\pi}\int d^2\boldsymbol{x}'\:\kappa(\boldsymbol{x}')\:\log|\boldsymbol{x} - \boldsymbol{x}'|.
\end{equation}

\begin{figure}[h]
\centering
    {\includegraphics[width=8cm,height=8cm]{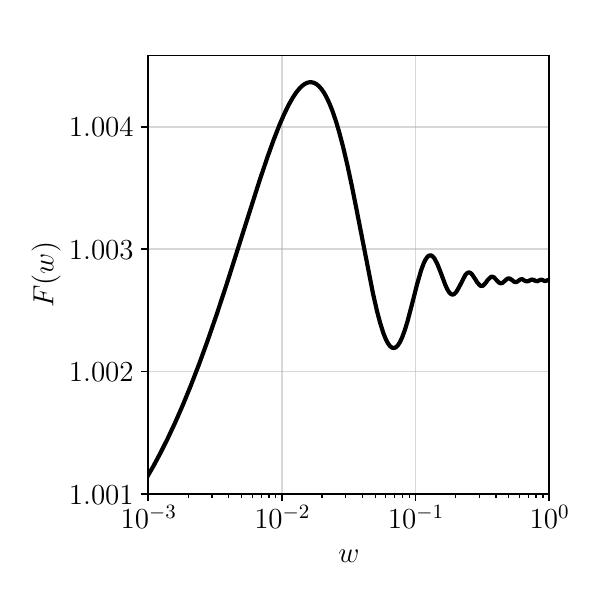}}
    \caption{The WO regime, marked by the onset of the first peak, followed by oscillatory behaviour due to the interference between the multiple images. The computation of $F(w)$ requires solving the complex integral in Eq.~\ref{diff-int-scaled}, which inherently depends on the lens' density distribution through the lensing potential.}
    \label{fig:NFW-FW}
\end{figure}

The dimensionless version of the time delay function (Fermat potential) takes the form
\begin{equation}
\label{phi-scaled}
    \phi(\boldsymbol{x},\boldsymbol{y}) = \frac{1}{2}|\boldsymbol{x}-\boldsymbol{y}|^2 - \psi(\boldsymbol{x}) - \phi_m(\boldsymbol{y}).
\end{equation}
Applying the scaling relation from Eq.~\ref{x-y-scaled}--\ref{psi-scaled} , the diffraction integral $F(f)$ can be expressed as 
\begin{equation}
    \label{diff-int-scaled}
    F(w) = \frac{w}{2 \pi i}\int d^2\boldsymbol{x} e^{iw\phi(\boldsymbol{x},\boldsymbol{y})}
\end{equation}
where $w$ is the dimensionless frequency
\begin{equation}
    \label{dimless-freq}
    w\equiv8\pi GM_{Lz}f
\end{equation} 
corresponding to an effective redshifted lens mass defined as
\begin{equation}
\label{mlz-theory}
    M_{Lz}\equiv\frac{\xi_0^2}{4Gd_\mathrm{eff}}.
\end{equation}
The Fourier transform of $ F(w)$ can be expressed as  $\mathcal{I}(\tau)$, as in Eq.~\ref{Itau-scaled}
\begin{equation}
    \label{Itau-scaled}
    \mathcal{I}(\tau) = \int_{-\infty}^{+\infty} \frac{iF(w)}{w} e^{-i w \tau} \, dw.
\end{equation}
We are interested in the WO regime, which is marked by the onset of the first peak, followed by oscillatory behaviour in the amplification factor. These oscillations are due to the interference between the multiple images (Fig.~\ref{fig:NFW-FW}). The computation of the amplification factor $F(w)$, requires solving the complex integral in Eq.~\ref{diff-int-scaled}, which inherently depends on the lens' density distribution through the lensing potential.

To accomplish this, we leverage the Gravitational Lensing of Waves (GLoW) code \cite{villarrubiarojo2024glownovelmethodswaveoptics}\footnote{\url{https://miguelzuma.github.io/GLoW_public/index.html}}, a {Python library} designed for wave optics calculations. The GLoW framework enables precise evaluation of $F(w)$ by implementing efficient numerical techniques tailored for lensing scenarios. For our analysis, we utilize the \texttt{SingleIntegral} method, a highly optimized approach written in \texttt{C}. This method is particularly well-suited for axisymmetric lensing configurations and handles both single and multiple image systems with high accuracy.

As described in \cite{villarrubiarojo2024glownovelmethodswaveoptics}, the \texttt{SingleIntegral} method is valid for generic but symmetric lenses. It involves two main steps: first, calculating the dimensionless diffraction integral $\mathcal{I}(\tau)$, and second, applying a direct Fast Fourier Transform (FFT) to obtain the amplification factor $F(w)$. This approach ensures computational efficiency while retaining the fidelity of WO phenomena, such as the oscillatory interference patterns resulting from image multiplicity. Additionally, the GLoW code provides flexibility for studying a wide range of lens models, including substructure, making it a versatile tool for studying WO features.

\section{Fuzzy dark matter}
\label{FDM}

CDM simulations have provided a successful framework for understanding the large-scale phenomena in the Universe, explaining the matter-energy relationship and the evolution of cosmic structures. However, these simulations face significant challenges in explaining galactic or sub-galactic scale observations \cite{Bullock_2017, Del_Popolo_2017, Bull_2016, WeinbergBullockGoverntro}. Some of these challenges are as follows --

\begin{itemize}
    \item \emph{Core cusp problem} -- Dark matter haloes predicted by CDM are expected to have a cuspy core. However, observations reveal a deviation from this prediction, with haloes demonstrating a more uniform density towards their centres \cite{Dubinski1991496, de_Blok_2009, Walker_2011}.
    \item \emph{Too big to fail} -- The predicted number of high luminosity satellites in CDM simulations exceeds the observed count \cite{Boylan_Kolchin_2012}. 
    \item \emph{Baryonic Tully-Fisher Relation} -- The tight relationship between the circular velocity of galaxies and their total baryonic mass demonstrates a distinct slope and reduced scatter compared to predictions from CDM and hierarchical structure formation theories \cite{McGaugh_2011}. 
    \item \emph{Missing Satellites} -- Prediction of over-abundance of halo substructures relative to the observed number of satellite galaxies \cite{Moore_1999, Klypin_1999}.
\end{itemize}

To address these challenges, an alternative dark matter model is proposed, consisting of extremely light bosons or axions with masses in the range, \( m_\phi \sim 10^{-22} - 10^{-23} \) eV. This model, known as ``Fuzzy Dark Matter'' (FDM) \cite{Schive_2014, schive2014cosmic, Hu_2000, Hui_2017, Mina_2022, Mocz_2019, Ferreira:2020fam, Hui:2021tkt}, replicates the behaviour of Cold Dark Matter (CDM) on large scales, with deviations becoming apparent only at smaller scales (Sec.~\ref{nature_FDM}). 

In the following section, we briefly explore the properties of FDM halos and their implications on cosmological scales (Sec.~\ref{nature_FDM}). Additionally, we discuss the proposed formation mechanisms of FDM halos (Sec.~\ref{form_FDM}) and review some astrophysical constraints (Sec.~\ref{const_FDM}) on the mass of FDM halos arising from their quantum nature. Finally, in Sec.~\ref{modelling-FDM}, we discuss how we model our FDM halo.

\subsection{Nature and consequences of fuzzy dark matter}
\label{nature_FDM}
On large scales, FDM effectively mimics the behaviour of CDM. However, on smaller scales, all the collapsed filaments are distinguished by the presence of a stable core, called \emph{soliton}. The core represents a dense structure of coherent standing waves, obeying the Schrödinger-Poisson equation. In every FDM halo, a flat density distribution towards the centre is facilitated by the existence of these cores. These cores continue to grow as additional matter is accreted to the expanding potential well. However, due to the quantum pressure exerted by FDM, an upper limit exists on the amount of infalling matter. Consequently, the core is surrounded by virialized halos. On small scales, FDM exhibits the presence of interference patterns which continuously fluctuate in density, creating a stable system by generating quantum and turbulent pressure to counterbalance the gravitational attraction of infalling material. 

FDM particles can be expressed in terms of de Broglie's wavelength as \cite{Hui_2017}
\begin{equation}
\label{de_brog_eq_FDM}
    \lambda = \frac{\hbar}{mv} = 1.92\:\bigg( \frac{10^{-22}\:\mathrm{eV}}{m_\phi} \bigg) \bigg(\frac{10\:\mathrm{km\:s}^{-1}}{v} \bigg)\mathrm{kpc}.
\end{equation}

The extremely small mass of FDM results in a very long de Broglie wavelength, giving rise to wave-like behaviour. In the early Universe (very high redshifts), when matter density was low, the wave-like nature of FDM particles was more pronounced. The interference patterns generated by the long de Broglie wavelength inhibit the collapse of matter into low-mass halos. Instead, it tends to smear out density fluctuations, preventing the formation of localized structures. The interference effects due to the large wavelengths hinder the clumping of matter into small-scale structures, making low-mass halos harder to form.

\subsection{Formation of fuzzy dark matter halo}
\label{form_FDM}
The evolution of the FDM halo involves several distinct stages that contribute to the formation of structures in the Universe. Beginning as small clumps within a filamentary structure, dark matter gradually accretes onto the growing potential wells, eventually leading to full virialization and the formation of the first halo. The gravitational influence plays a crucial role in initiating the collapse of the scalar field associated with FDM. On scales comparable to the de Broglie wavelength, the gravitational attraction overcomes the quantum pressure, initiating the collapse process. During the collapse, the energy of the scalar field redistributes away from the region of the highest density. This redistribution of energy results in the formation of a stable solitonic core within the halo. Once a stable core has formed, further accretion of matter onto the core is not possible. During these stages, the quantum nature of dark matter becomes apparent through the development of interference patterns. These patterns provide evidence of the wave-like behaviour inherent to FDM particles.

\subsection{Constraints on fuzzy dark matter halo}
\label{const_FDM}

We can constrain the minimum allowed mass for the FDM halo using the Jeans mass stability criterion \cite{Hui_2017, Mina_2022}. This follows as 
\begin{equation}
    M > 1.47\times10^7(1 + z)^{3/4}\left(\frac{\Omega_\mathrm{FDM}h^2}{0.12}\right)^{1/4}\left(\frac{m_\phi}{10^{-22}\mathrm{eV}}\right)^{-3/2}\:\mathrm{M_\odot}
\end{equation}
where $h\equiv H_0/100$km/s/Mpc is the dimensionless reduced Hubble constant and $\Omega_\mathrm{FDM}$ represents the current fraction of the universe's critical energy density contributed by the dark matter component. 

Although a lower bound exists for the FDM halo mass, if the maximum mass is significantly larger than the soliton mass, the soliton constitutes only a small fraction of the total halo mass. This suppresses the quantum nature of the halo, consequently, the WO features of the FDM halo resemble those of a CDM halo (see Fig.~\ref{vary-y-FDM-I} and Fig.~\ref{vary-y-FDM-F}, top to bottom row).

Using the quantum nature of the FDM halo, it is thus possible to provide an upper bound on the central density of the FDM halo. Restricting the de Broglie's wavelength, which cannot exceed the virial radius, $R_\mathrm{vir}\sim GM/v^2$ of an equilibrium self-gravitating system of mass $M$ and the boson mass $m_\phi$, implies $r \geq  {\hbar^2/(GMm_\phi^2)}$. Then the central density $\rho_c$ (Eq.~\ref{rho_c_soliton} \cite{Hui_2017}) follows
\begin{equation}
\label{rho_c_soliton}
    \rho_c \leq 7.051\times10^9 \bigg(\frac{m_\phi}{10^{-22}\mathrm{eV}}\bigg)^6 \left(\frac{M}{10^9\mathrm{M}_\odot}\right)^4\:\mathrm{M_\odot\:kpc}^{-3}.
\end{equation}

\subsection{Modelling the fuzzy dark matter halo}
\label{modelling-FDM}

\begin{figure*}
    \begin{subfigure}[]{\includegraphics[width=8cm,height=8cm]{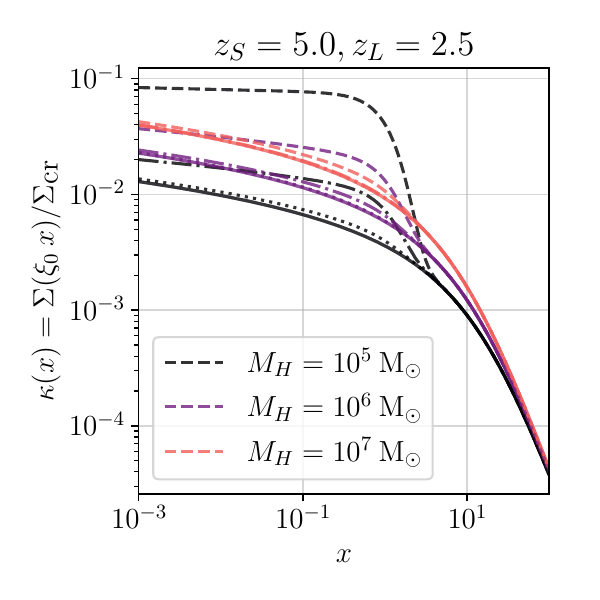}}\end{subfigure}
    \begin{subfigure}[]{\includegraphics[width=8cm,height=8cm]{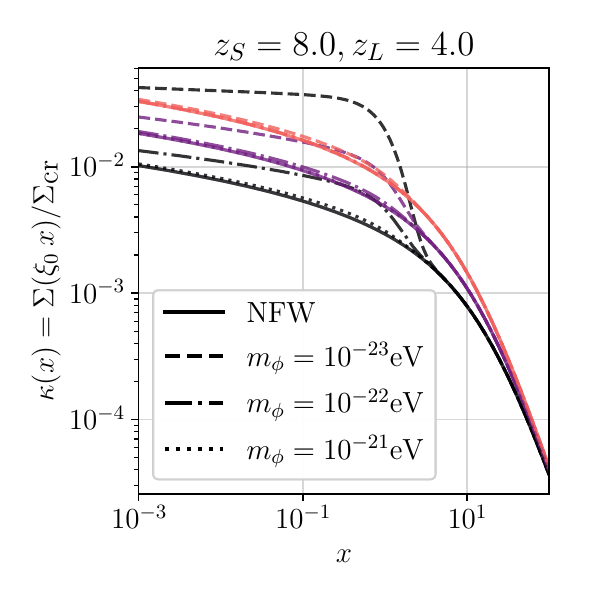}}\end{subfigure}
    \begin{subfigure}[]{\includegraphics[width=8cm,height=8cm]{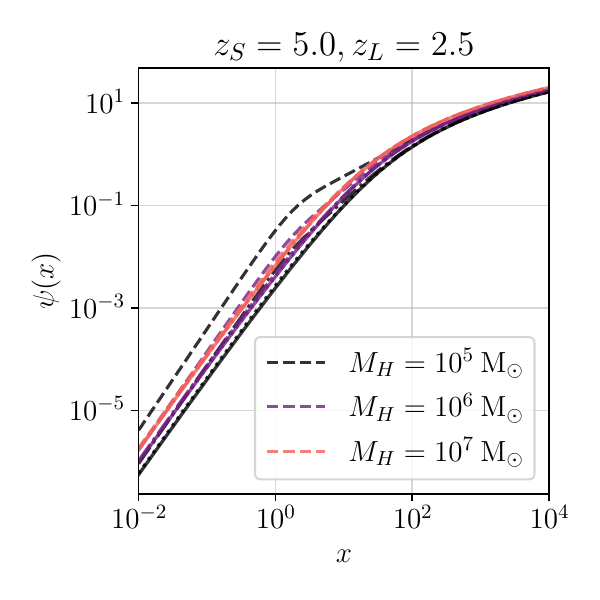}}\end{subfigure}
    \begin{subfigure}[]{\includegraphics[width=8cm,height=8cm]{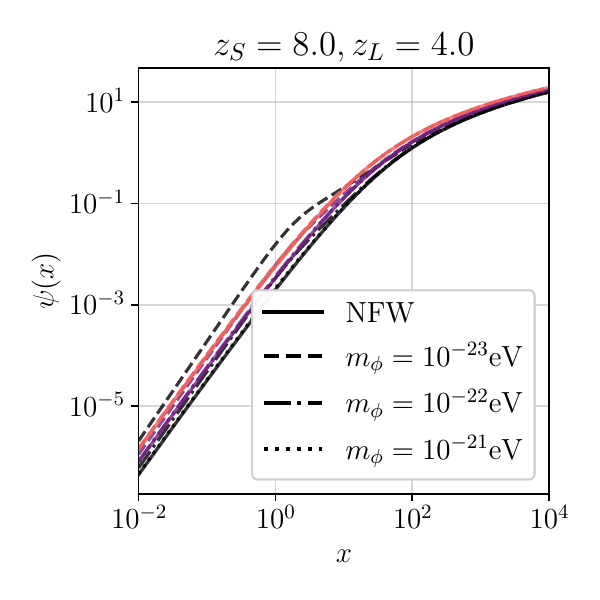}}\end{subfigure}
    \caption{Normalized projected density $\Sigma(\boldsymbol{\xi})$ (top row) and Lensing potential $\psi(\boldsymbol{x})$ (bottom row), for FDM and NFW (solid). The results are presented for three halo masses, $M_\mathrm{H} \in  {\{10^5,\:10^{6},\:10^{7}\}}\:\mathrm{M}_\odot$ each varying for two boson mass $m_\phi \in  \{10^{-21},\:10^{-22}\:10^{-23}\}$ eV (dashed, dot-dashed and dotted respectively). These quantities are calculated for fixed lens redshifts of $z_L =2.5$ {(left column)} and $z_L =4.0$ {(right column)}.}
\label{sigma_psi_comp_FDM_NFW}
\end{figure*}

We model the FDM density distribution by incorporating a solitonic core in an NFW halo. With deviations from NFW only at small scales, the cusp is replaced with a solitonic core. The complete profile is modelled as 
\begin{equation}
    \rho_{\mathrm{FDM}}(r) = \Theta(r_\epsilon-r)\rho_{\mathrm{SOL}}(r) + \Theta(r-r_\epsilon)\rho_{\mathrm{NFW}}(r).
    \label{FDM-eq}
\end{equation}
where $\Theta$ represents the Heaviside step function, and $r_\epsilon$ is the transition radius from solitonic core to NFW profile \footnote{$r_\epsilon : \rho_{\mathrm{SOL}}(r_\epsilon) = \rho_{\mathrm{NFW}(r_\epsilon})$}.

\begin{equation}
    \label{SC-eq}
    \rho_{\mathrm{SOL}}(r) = 1.9\times10^9\:\left(\frac{h}{a}\right)\left(\frac{m_\phi}{10^{-23}\mathrm{eV}}\right)^{-2}\left(\frac{R_c}{\mathrm{kpc/h}}\right)^{-4}\left(1+9.1\times10^{-2}\frac{r^2}{R_c^2}\right)^{-8}\:\frac{\mathrm{M}_\odot}{(\mathrm{kpc/h})^3}
\end{equation}

\begin{equation}
\label{Rc}
    R_c = 1.6 ha^{1/2}\left(\frac{m_\phi}{10^{-22}\mathrm{eV}}\right)^{-1} \left(\frac{\zeta(z_L)}{\zeta(0)}\right)^{-1/6} \left(\frac{\mathrm{M_{vir}}}{10^9\textrm{M}_\odot}\right)^{-1/3}\:\mathrm{\frac{kpc}{h}}
\end{equation}
\begin{equation}
\label{Mc}
    M_c =  \frac{1}{4} a^{-1/2}\left(\frac{\zeta(z_L)}{\zeta(0)}\right)^{1/6} \left(\frac{\mathrm{M_{vir}}}{M_\mathrm{min,0}}\right)^{1/3} M_\mathrm{min,0}
\end{equation}
\begin{equation}
\rho_{\mathrm{NFW}}(r) = \rho_{0} \left(\frac{r}{R_s}\right)^{-1} \left(1 + \frac{r}{R_s}\right)^{-2} 
    \label{NFW-eq}
\end{equation}

Following \cite{schive2014cosmic}, the density profile of the solitonic core can be described as in Eq.~\ref{SC-eq}, where $R_c$ represents the core radius. The boson mass has a strong influence on the core radius, which can be expressed as in Eq.~\ref{Rc}. From \cite{Schive_2014} mass of the core $M_c$ can be expressed in terms of $\mathrm{M_{vir}}$ as in Eq.~\ref{Mc}, where $\zeta(z_L)\equiv 18\pi^2+82(w_\mathrm{m}-1)-39(w_\mathrm{m}-1)^2 w_\mathrm{m}^{-1}\sim 350 (180)$ for $z_L=0\:(z_L\geq 0)$ \cite{bryan1998statistical} and $M_\mathrm{min}\equiv375^{-1/4}32\pi \zeta(0)^{1/4}\rho_c(H_0m_\phi/\hbar)^{-3/2}w_\mathrm{m}^{-3/4}\sim 4.4\times10^7(m_\phi/10^{-22}\mathrm{eV})^{-3/2}\:\mathrm{M}_\odot$.

NFW profile can be expressed as in Eq.~\ref{NFW-eq}, where $R_s={\mathrm{R_{vir}}}c_{\mathrm{NFW}}^{-1}$
represents the transition between a log-slope $-1$ in the central part and a log-slope of $-3$ in the outer, and $c_{\mathrm{NFW}}$ is the concentration parameter. The central density $\rho_0$, can be expressed in terms of $c_{\mathrm{NFW}}$ as 

\begin{equation}
    \rho_0 = \frac{\rho_c\Delta_cc_{\mathrm{NFW}}^3}{3\left[\log(1+c_{\mathrm{NFW}})-c_{\mathrm{NFW}}(1+c_{\mathrm{NFW}})^{-1}\right]}
\label{eq_rho_c-rho_0}
\end{equation}
where, $\rho_c$ represents the critical density and $\Delta_c$ is the overdensity. 

In Fig.~\ref{sigma_psi_comp_FDM_NFW} we show the convergence, $\kappa = \Sigma/\Sigma_{\textrm{cr}}$ for FDM (dashed, dotted-dashed) and NFW (solid) for a range of halo and boson mass at a fixed redshift. From Fig.~\ref{sigma_psi_comp_FDM_NFW}, we can see that low-mass halos with lighter boson mass $m_\phi$ show a larger deviation from NFW-only halos. This effect becomes even more pronounced in low-redshift halos, where the deviation from NFW-only profiles is significantly enhanced.

\section{Wave optics features of FDM}
\label{wave_optics_FDM}

\begin{figure*}
\centering
    \begin{subfigure}[]{\includegraphics[width=7cm,height=19cm]{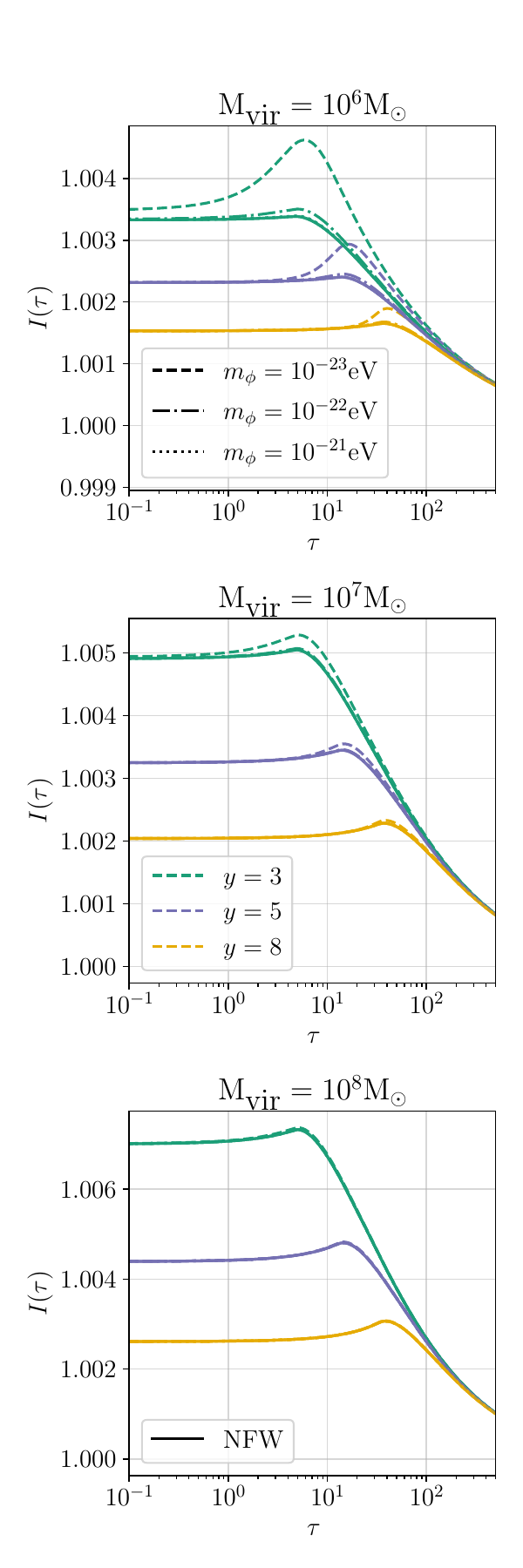}}\end{subfigure}
    \begin{subfigure}[]{\includegraphics[width=7cm,height=19cm]{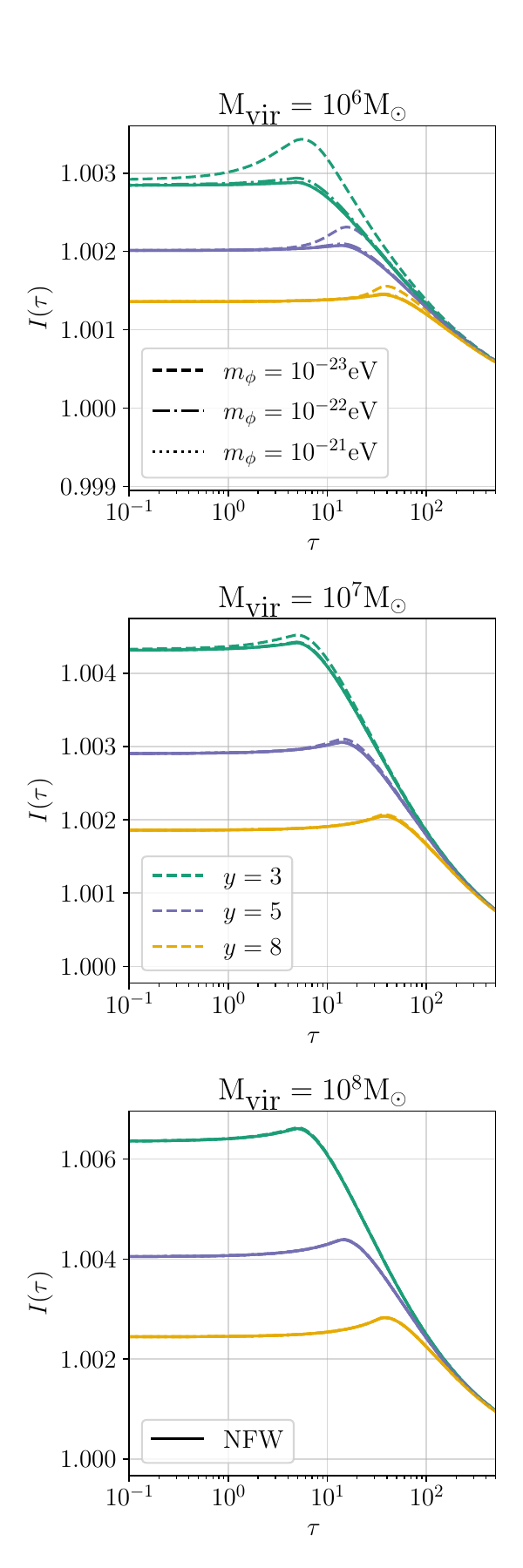}}\end{subfigure}
    \caption{Wave Optics (WO) features for NFW (solid) and FDM (dashed, dotted-dashed and dotted). $\mathcal{I}(\tau)$ is shown for the impact parameter $y \in \{3,5 ,8\}$ for different $m_\phi \in \{10^{-23}, 10^{-22}, 10^{-21}\}$ eV, each varying over three virial masses M$_\textrm{vir} \in \{10^6, 10^7, 10^8\}\:\textrm{M}_\odot$ with a fixed lens redshift of $z_L=2.5$ (left column) and $z_L=4.0$ (right column).}
    \label{vary-y-FDM-I}
\end{figure*}

\begin{figure*}
\centering
    \begin{subfigure}[]{\includegraphics[width=7cm,height=19cm]{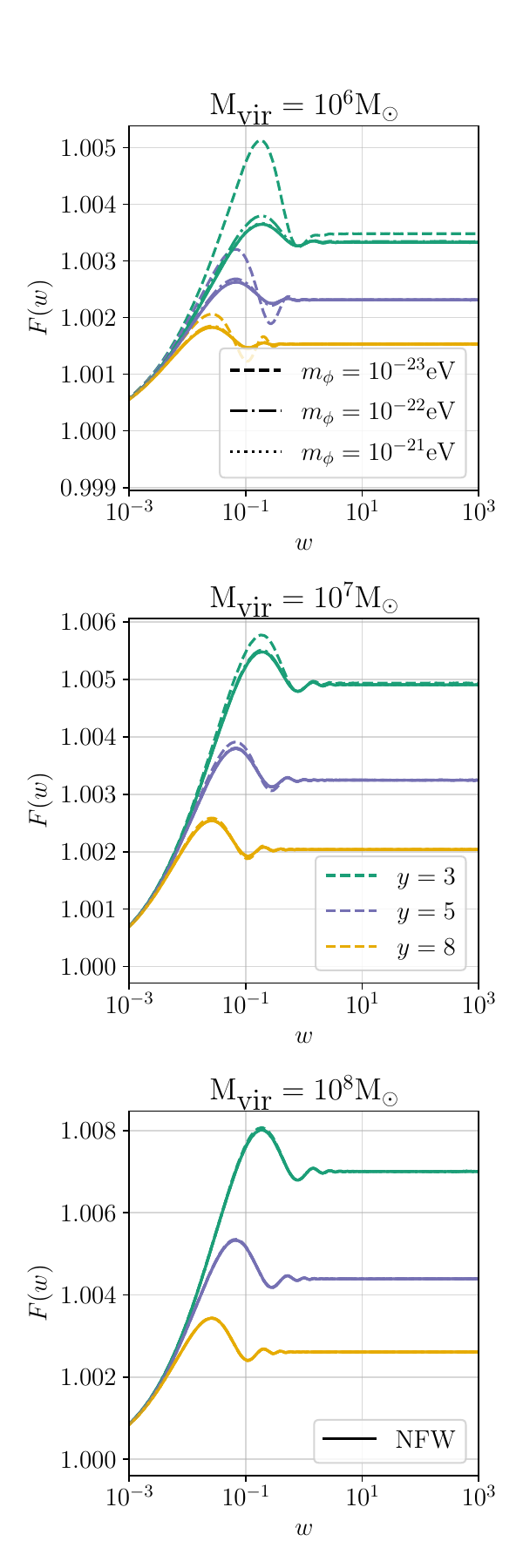}}\end{subfigure}
    \begin{subfigure}[]{\includegraphics[width=7cm,height=19cm]{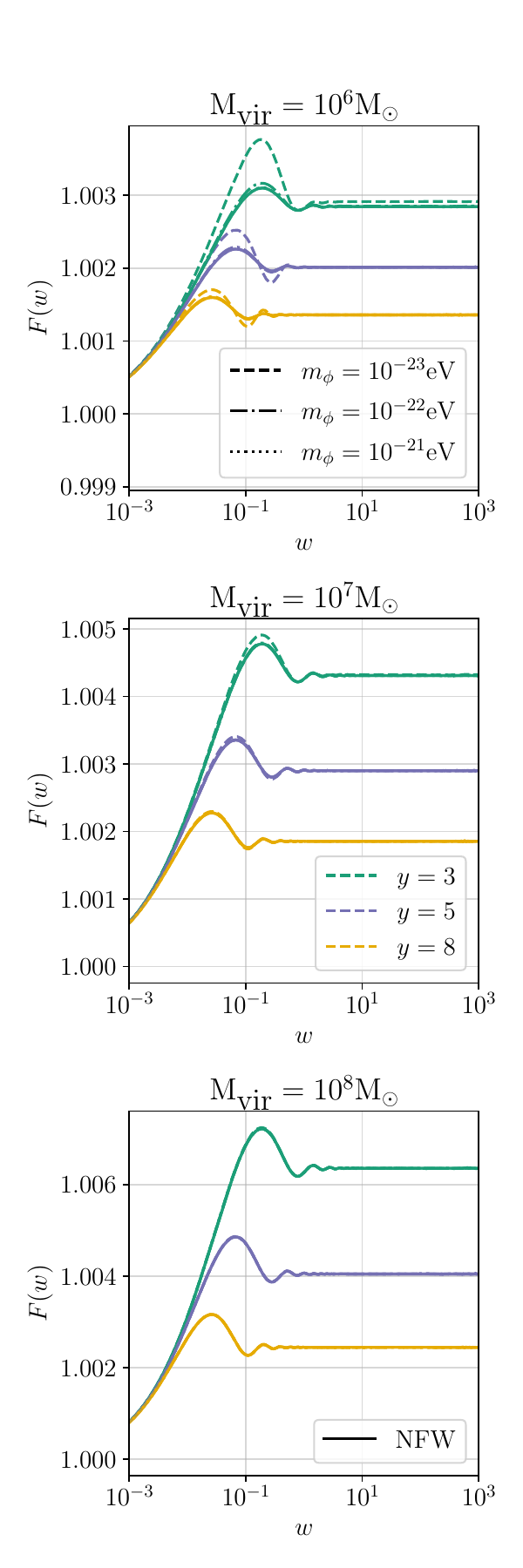}}\end{subfigure}
    \caption{(Continued) WO features -- $F(w)$ with the same parameters as in the previous figure, Fig.~\ref{vary-y-FDM-I}.}
    \label{vary-y-FDM-F}
\end{figure*}

With the FDM halo now modelled, we can shift our focus to its wave optics (WO) features. Following the procedure outlined in Sec.~\ref{theory}, we can compute the projected density $\Sigma_\textrm{FDM}$, using Eq.~\ref{sigma} and consequently the lensing potential $\psi(\boldsymbol{x})$.

As described in Sec.~\ref{modelling-FDM}, we model the FDM halo as a solitonic core (SC) embedded within an NFW halo. 
To compute the lensing potential for this composite halo, it is essential to {rescale the quantities associated with SC to be able to embed in the NFW halo. Based on our modeling approach, we show this in two steps. In Appendix~\ref{xi_0_eval} we calculate the $\xi^\mathrm{SC}_0$, which is the characteristic length (Eq.~\ref{x-y-scaled}) and then rescale the associated quantities of SC in Appendix~\ref{rescaling_SC}.}
With {the rescaled quantities} in hand, we proceed to evaluate the lensing features produced by the FDM halo and compare them with those of a pure NFW halo (see Fig.~\ref{vary-y-FDM-I} and Fig.~\ref{vary-y-FDM-F}). Using the computational method outlined in \cite{Savastano_2023} and \cite{villarrubiarojo2024glownovelmethodswaveoptics}, we examine the evolution of the time-domain integral $\mathcal{I}(\tau)$ (Eq.~\ref{Itau-scaled}) as well as the amplification factor $F(w)$ (Eq.~\ref{diff-int-scaled}).

The results are presented across a range of parameters. We show the WO features for a source with impact parameters $y \in \{3, 5, 8\}$. For comparison, we plot these features for three boson masses, $m_\phi \in \{10^{-23}, 10^{-22}, 10^{-21}\}$ eV, across three halos with virial masses $\mathrm{M_{vir}} \in \{10^6, 10^7, 10^8\} \:\mathrm{M_\odot}$. Additionally, we include the same features for an NFW-only halo (solid lines), while fixing the lens redshift at $z_L = 2.5$ and $z_L = 4.0$.

In the following subsections, we focus on the features present in the time-domain integral $\mathcal{I}(\tau)$ and the amplification factor $F(w)$ for the FDM lens, and how these differ from those of the NFW lens.

\subsection{Time domain integral : $\mathcal{I}(\tau)$}
Both FDM and NFW lenses exhibit characteristic behaviour in $\mathcal{I}(\tau)$, which first attains a maximum at $\tau_c =\phi(\boldsymbol{x} = 0, \boldsymbol{y})$ and then gradually decreases to approach unity. The peak is due to the high curvature in the constant $\phi$ contours as they approach the lens. The peak's height, width, and position depend on the lens matter distribution and the impact parameter.  In Fig.~\ref{vary-y-FDM-I}, the peak of $\mathcal{I}(\tau)$ shifts to higher $\tau$ with decreasing height as $\boldsymbol{y}$ increases. This occurs because an increase in $\boldsymbol{y}$ results in the interaction of higher $\tau$ with the lens centre, forcing the peak to move towards a larger value of $\tau$. The peak's height can be interpreted within the non-perturbative single-image framework introduced in \cite{Savastano_2023}, represented by Eq.~\ref{Itau_savastano2023weakly}

\begin{equation} \label{Itau_savastano2023weakly} \mathcal{I}(\tau) = \int^{2\pi}_0 d\theta \frac{R(\theta)}{|\partial_R\phi|}. \end{equation}

{In this context, $\theta$ denotes the polar angle in a coordinate system centered at $\boldsymbol{x}_m$, where $\boldsymbol{x}_m$ is the minimum of the Fermat potential such that $\phi(\boldsymbol{x}_m) = 0$. The spatial coordinates $(x_1, x_2)$ are then expressed as $x_1 = x_{m,1} + R\cos{\theta}$ and $x_2 = x_{m,2} + R\sin{\theta}$, where $R$ is the radial distance from $\boldsymbol{x}_m$ and $\theta$ is the polar angle (see~\cite[Sec.~II.B]{Savastano_2023} for details regarding the notation).}

We also observe that as M$_\textrm{vir}$ increases (from top to bottom in Fig.~\ref{vary-y-FDM-I}), the peak position of $\mathcal{I}(\tau)$ remains unchanged for the reasons described above. However, the height and width are affected, reflecting differences in the lens matter distribution. More massive halos (larger M$_\textrm{vir}$) exhibit deeper gravitational potentials, resulting in taller and narrower peaks (as described in Eq.~\ref{Itau_savastano2023weakly}). 

While comparing these features with NFW, FDM lenses show deviation for low $m_\phi$, which are particularly pronounced for low M$_\textrm{vir}$. The deviations diminish as $m_\phi$ and M$_\textrm{vir}$ increases. For large $m_\phi$, the de Broglie's wavelength becomes small (Eq.~\ref{de_brog_eq_FDM}), leading to clumping of large amounts of matter (Sec.~\ref{form_FDM}). This results in a small soliton radius (Eq.~\ref{Rc}) and the suppression of distinctive FDM features. Consequently, for large M$_\textrm{vir}$, FDM and NFW exhibit similar characteristics. 

The features are consistent across redshifts, but slightly higher values of $\mathcal{I}(\tau)$ are observed at lower redshifts for low M$_\textrm{vir}$. This arises from longer accretion cycles at lower redshifts, leading to a heavier core for the same M$_\textrm{vir}$. However, for larger M$_\textrm{vir}$ or $m_\phi$, this effect diminishes due to the reduced core-to-halo size ratio.

\subsection{Amplification factor : $F(w)$}
 
The presence of peaks in $\mathcal{I}(\tau)$ corresponds to the oscillatory behaviour observed in $F(w)$. From Fig.~\ref{vary-y-FDM-F}, a similar shift in the position of the first peak in $F(w)$ is evident, mirroring the behaviour of $\mathcal{I}(\tau)$. Sharper peaks in $\mathcal{I}(\tau)$ lead to more pronounced features in $F(w)$  and slower-decaying oscillations that persist at larger values of $w$. These oscillations in $F(w)$ arise from the varying phase differences between multiple images arriving via different paths.

Lower values of $m_\phi$ result in more prominent oscillations, attributable to the soliton core embedded within the FDM halo. Smaller $m_\phi$ values correspond to larger soliton cores, enhancing the oscillatory behaviour -- a feature also observed in cored lenses studied in \cite{tambalo2022gravitational}.

\section{Detection prospects}
\label{detection}
We will now address the prospects of observing these WO features of the FDM lens. First, we show how these features can help distinguish FDM halos from NFW halos. Then, we explore whether the FDM solitonic core enhances the detectability of weakly lensed signals.

\subsection{Identifiability of solitonic cores}

We aim to determine whether NFW and FDM halos can be distinguished by their WO lensing imprints. To do this, we compare the amplification factor $F(w)$ for both profiles.

\begin{figure}[h]
\centering
    {\includegraphics[width=8cm,height=8cm]{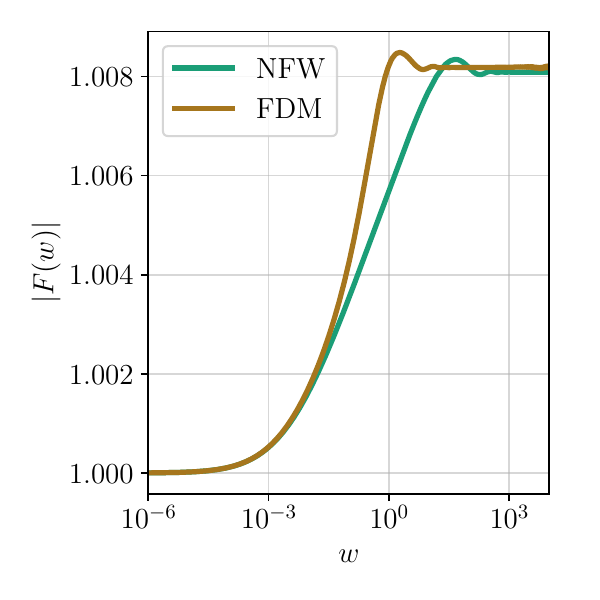}}
    \caption{Comparison between the amplification factor $F(w)$, for an NFW and an FDM halo {of virial mass M$_\textrm{vir}=10^6~\mathrm{M}_\odot$ at source redshift, $z_S=8.0$ and lens at $z_L=4.0$ for a boson mass of $m_\phi=10^{-23}~$eV}, highlighting their differences across frequency $w$. The FDM impact parameter has been chosen to minimize the difference in amplification factors (see text). {For this setup, we have used $y_\mathrm{FDM} = 1.0$ and minimized $y_\mathrm{NFW}$, with an optimal value of 0.19}. This approach allows for a direct comparison of the two profiles under consistent observational conditions, illustrating the distinguishability of the halo models.}
    \label{fig:opt-FDM-NFW}
\end{figure}

To ensure a meaningful comparison, we must account for potential differences in the (unknown) lens parameters. Thus, we fix the impact parameter for the FDM profile $y_{\textrm{FDM}}$and vary $y_{\textrm{NFW}}$ for the NFW profile. This approach is similar to the one discussed in Ref.~\cite{GilChoi:2023ahp}. The optimization problem can be mathematically expressed as:
\begin{equation}
\label{eq:optimization}
\min_{y_{\text{NFW}}} \int \left| F_{\text{NFW}}(w, y_{\text{NFW}}) - F_{\text{FDM}}(w, y_{\text{FDM}}) \right| \, dw.    
\end{equation}
Here, the integral is taken over the dimensionless frequency range of interest and could be weighted to account for the sensitivity of observations (e.g.~including signal strength and detector sensitivity). For simplicity, we will perform the optimization as Eq.~\ref{eq:optimization} in a range of $w$ that captures both the onset of diffraction, where $F \sim 1$, and the geometric optics limit, where $F \to \text{const}$ (see Fig.~\ref{fig:opt-FDM-NFW}). This approach allows us to determine the optimal value of $y_{\textrm{NFW}}$, which minimizes the discrepancy between the amplification factors of the two models.

Fig.~\ref{fig:opt-FDM-NFW} shows how WO features can distinguish between the NFW and FDM, even when allowing for different impact parameters in each lens type. The amplification can be distinguished clearly at intermediate frequencies, even when the optimized impact parameter differs by a factor $\sim 5$.
Our results are also robust to varying the halo mass in both profiles (i.e.~fixed $\mathrm{M_{vir}}$ for FDM but varied for NFW): because the dimensionless frequency $w$ scales with a monotonic function of the mass, changing the $\mathrm{M_{vir}}$ in the FDM profile amounts to a horizontal shift of the curve shown in Fig.~\ref{fig:opt-FDM-NFW}. This shift can slightly reduce the disagreement between both predictions, it only affects a narrow range of frequencies.

{For the single-image regime, the distinguishability of profiles relies on WO effects, via contributions from all regions of the lens plane in Eq.~(\ref{ampl-fac}). In contrast, in the GO limit, only the signal's amplitude is modified, which is degenerate with a change in the impact parameter and the source's luminosity distance.}
Note also that WO lensing features are not strongly degenerate with intrinsic source parameters, at least for non-spinning binaries~\cite[Fig.~7]{Brando_2025}. We aim to explore the distinguishability of solitonic cores in more detail in future works.

\subsection{Enhanced detection probability?}

We now address whether a dense solitonic core increases the probability of detecting WO lensing by FDM halos. This is plausible, as more concentrated profiles typically enhance detection prospects: the effect has been shown explicitly for NFW~\cite[Fig.~15]{Brando_2025}, as well as indirectly by comparing NFW with singular-isothermal profiles~\cite{Fairbairn_2023,Savastano_2023,Caliskan:2023zqm,Guo:2022dre}.

\begin{figure*}
    \includegraphics[height=7cm, width=15cm]{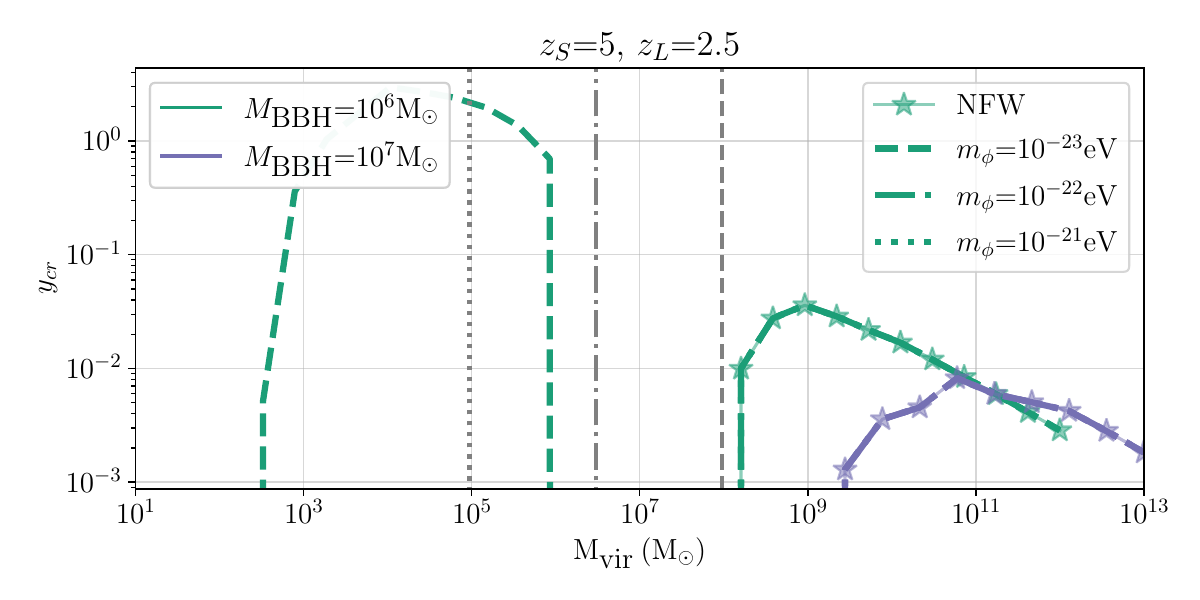}
    \caption{$y_\textrm{cr}$ for NFW and FDM for a lensed GW signal from a merger of an equal massive binary black hole of $10^6\textrm{M}_\odot$ (green) and $10^7\textrm{M}_\odot$ (purple). The source/lens redshifts are fixed $z_S=5$ and $z_L=2.5$. Lines show predictions for NFW (solid, stars) and three different boson masses (dashed-dotted from lighter to heavier). The vertical lines correspond to the lowest allowed halo masses for $w_{\rm FDM}=0.3$. FDM does not increase the detectability of isolated halos with respect to NFW: the boost for light halos $\rm{M_{vir}}<10^6M_\odot$ corresponds to objects that can not form in the ultra-light boson mass range.}
\label{fig:y_crit}
\end{figure*}

We will compute the highest impact parameter at which WO features can be potentially detected, $y_{\rm cr}$ following the method described in Ref.~\cite{Savastano_2023}. The critical impact parameter offers a good characterization of lensing probabilities, which scale as the enclosed surface $\propto y_{\rm cr}^2$. 
If $y_{\rm cr}$ deviates substantially from NFW, detection probabilities can be obtained by accounting for the halo mass function to obtain the (optical depth)~\cite{Savastano_2023,Caliskan:2023zqm,Brando_2025}.

The detectability of the lensing features can be studied using \emph{mismatch}, which is expressed as 
\begin{equation}
    \label{mismatch}
    \mathcal{M} = 1-\frac{\langle h_L|h_0 \rangle}{\sqrt{\langle h_L|h_L \rangle \langle h_0|h_0 \rangle}}
\end{equation}
where symbols have their usual meaning. 
Eq.~\ref{mismatch} depends on the normalized noise-weighted inner product between the lensed GW signal ($h_L$) and unlensed GW signal ($h_0$), which is defined as 
\begin{equation}
    \langle h_L|h_0 \rangle = 4\:\mathrm{Re}\:\int_0^{\infty}\frac{df}{S_n(f)}\tilde h_L(f) \tilde h_0(f)
\end{equation}
where $\tilde h_L(f)$  is the frequency domain waveform and $S_n(f)$ represents the sky-averaged one-sided detector power spectral density (PSD). The signal-to-noise ratio SNR can now be defined in terms of this noise-weighted inner product as SNR $\equiv \sqrt{\langle h|h \rangle}$. We follow the Lindblom criterion \cite{PhysRevD.78.124020}, which states that the distinguishability condition is fulfilled only if $\langle \delta h|\delta h \rangle < 1$ where $\delta h \equiv h_L - h_0$. For comparable SNR signals, which would be a case for weakly lensed and unlensed GW signals, the criteria require $\mathcal{M}\times \mathrm{SNR}^2>1$ \cite{Savastano_2023}.

For our analysis, we fix the GW source to be an equal mass massive black hole. For simplicity, we have excluded the spin parameter and inclination and polarization effects are sky-averaged. We have used IMRPhenomXHM \cite{Garc_a_Quir_s_2020}, which is a waveform for non-precessing black-hole binaries to generate the GW waveform.

The condition $\mathcal{M}\times \mathrm{SNR}^2=1$ marks the region of critical impact parameter $y_\mathrm{cr}$, which is the minimum value of $y$ required for the detectability of WO features.  In Fig.~\ref{fig:y_crit}, we show the critical impact parameter as a function of the halo mass, for both NFW and FDM profiles. We consider two (total) masses for the binary-black hole and three cases of ultra-light boson $m_\phi \in \{10^{-23}, 10^{-22}, 10^{-21}\}$ eV. For high mass halos, $\rm{M_{vir}}\geq 10^8M_\odot$ the critical impact parameter of FDM is almost identical to that of NFW. Only for the lightest boson particles ($10^{-23}$eV) do we observe a strong enhancement of $y_{\rm cr}$ at very low halo masses $\rm{M_{vir}}\sim 10^3-10^6M_\odot$. However, this is artificial, as such light halos are prevented from forming dynamically in ultra-light FDM models.

We therefore conclude that solitonic cores, although in principle distinguishable via WO features, do not increase the detectability of isolated halos with respect to the NFW model. This means that the rates will be comparable to those derived in Ref.~\cite{Brando_2025} (at least as long as the halo mass functions of FDM and CDM are close enough).  The assumption of isolated halos is an important caveat to this conclusion: more complex lens configurations can enhance the detectability of WO distortions, e.g.~in a halo+subhalo system~\cite[Fig.~17]{Brando_2025}.

\section{Conclusions}
\label{conclusion}
In this work, we investigated the prospects for detecting wave optics (WO) features in lensed gravitational wave (GW) signals from massive black hole mergers, lensed by dark matter halos composed of fuzzy dark matter (FDM). This focus is motivated by FDM's capacity to address observational challenges at galactic and sub-galactic scales, distinguishing it from the standard cold dark matter (CDM) paradigm \cite{Bullock_2017, Del_Popolo_2017, Bull_2016}. We focussed on the capability of future space-based GW detector Laser Interferometer Space Antenna (LISA) \cite{colpi2024lisadefinitionstudyreport} to probe unique WO features of the FDM halo.

Using semi-analytic modelling of FDM halo (Sec.~\ref{modelling-FDM}) and numerical computations (Sec.~\ref{theory}), we analysed the induced WO effects on GW signals. The presence of a solitonic core in FDM halos induces distinct oscillatory features in $F(w)$ which are absent in CDM lensing scenarios. While both FDM and CDM halos produce WO effects, the solitonic core in FDM introduces sharper peaks in $\mathcal{I}(\tau)$ and slower decaying oscillations in $F(w)$. This distinction provides a potential for differentiating between the two dark matter models using GW lensing observations. These features become more pronounced for lower boson masses $m_\phi$, since the core size increases with decreasing $m_\phi$, amplifying the effects due to the presence of a solitonic core.

The detectability of WO features depends strongly on the boson mass $m_\phi$ and the halo's virial mass $\mathrm{M_{vir}}$. LISA, with sensitivity to low-frequency GWs, is well-suited to detecting WO effects from FDM lenses. Our calculations suggest that for sufficiently massive and dense lenses, the unique soliton-induced features could serve as a direct probe of FDM's viability as a dark matter candidate. The main conclusions can be summarized as:
\begin{itemize}
    \item The increase in central density in FDM halos enhances the detectability of wave-optics features with respect to  NFW halos, expected in CDM theories (see Fig.~\ref{vary-y-FDM-F}).
    \item Precise observations can distinguish the FDM and NFW profile (for a given boson mass): the difference can not be masked as a different halo mass or impact parameter, see Fig.~\ref{fig:opt-FDM-NFW}.
    \item Despite leading to distinct wave-optics signatures, solitonic cores do not enhance the detectability of isolated halos with large offset from the source, see Fig.~\ref{fig:y_crit}. 
\end{itemize}

These results, however, have a few limitations that require further investigation. 
 Most importantly, we have restricted our analysis to symmetric and isolated halos for simplicity. However, it has been shown that more complex matter distributions enhance wave-optics phenomena -- this has been shown for external potentials \cite{Dai:2018enj}, in two-halo systems \cite[Fig.~17]{Brando_2025} and high densities of compact lenses~\cite[Fig.~3]{Zumalacarregui:2024ocb}.
Our semi-analytical model of FDM halo has been studied in the most idealized condition, neglecting any environmental factors, such as the effect of baryonic matter \cite{Chan_2015,Duffy_2010} or dark matter halo interactions \cite{painter2024attractivemodelsimulatingfuzzy}. Moreover, extracting the subtleties of WO features necessitates the development of parameter estimation pipelines accounting for deviations due to FDM~\cite{Cheung:2024ugg} adapted to LISA \cite{baghi2022lisadatachallenges}. These limitations will be addressed in future work.

Once lensed GWs are discovered, wave-optics features can be used to characterize the dark matter distribution and distinguish the central profile, e.g. the mass or density of the solitonic core, as well as the spin of the ultra-light boson (here a spin-0 field)~\cite{Lopez-Sanchez:2025osk}. Given the sensitivity of upcoming GW observatories on the ground~\cite{ET:2019dnz,Evans:2021gyd,Kalogera:2021bya} and in space~\cite{colpi2024lisadefinitionstudyreport,TianQin:2015yph,LISACosmologyWorkingGroup:2022jok,Sesana:2019vho}, wave-optics lensing may provide a long-sought breakthrough into the mystery of dark matter.

\acknowledgments
We express our sincere gratitude to G. Tambalo and H. Villarrubia-Rojo for their insightful discussions. S. Singh extends special thanks to G. Tambalo for his guidance in understanding the nuances of prior works and to H. Villarrubia-Rojo for his invaluable support with the GLoW Python library. Furthermore, S. Singh acknowledges the financial support provided by the ERASMUS+ mobility grant and extends gratitude to the AEI Potsdam for their assistance during the research stay.

\appendix

\section{Calculation of {$\xi^\mathrm{SC}_0$}}
\label{xi_0_eval}
{Here we show the calculation of $\xi^\mathrm{SC}_0$ which is the characteristic length (Eq.~\ref{x-y-scaled}) for Solitonic Core (SC).}

From Eq.~\ref{SC-eq}, we can integrate the matter distribution along the line of sight to obtain the projected surface density:
\begin{align}
    \Sigma^{\textrm{SC}} &= \int_{-\infty}^{\infty} 1.9\times10^9 ha^{-1}\left(\frac{m_\phi}{10^{-23}\mathrm{eV}}\right)^{-2}\left(\frac{R_c}{\mathrm{kpc/h}}\right)^{-3}\left(1+9.1\times10^{-2}\frac{r^2}{R_c^2}\right)^{-8} \textrm{d}\mathcal{Z} \nonumber \\ &=1.9\times10^9 ha^{-1}\left(\frac{m_\phi}{10^{-23}\mathrm{eV}}\right)^{-2}\left(\frac{R_c}{\mathrm{kpc/h}}\right)^{-3}\left(\frac{429 \pi}{2048 \times0.301}\right)\left(1+9.1\times10^{-2}\frac{\boldsymbol{\xi}^2}{R_c^2}\right)^{-15/2} \nonumber \\
    &= \Lambda\left(1+9.1\times10^{-2}\frac{\boldsymbol{\xi}^2}{R_c^2}\right)^{-15/2}{\frac{\mathrm{M}_\odot}{(\mathrm{kpc/h})^2}},
\end{align}
where  
\begin{equation}\label{eq:Lambda}
\Lambda = 1.9\times10^9 ha^{-1}\left(\frac{m_\phi}{10^{-23}\mathrm{eV}}\right)^{-2}\left(\frac{R_c}{\mathrm{kpc/h}}\right)^{-3}\left(\frac{429 \pi}{2048 \times0.301}\right).
\end{equation}
With (\ref{eq:Lambda}) we can express the lensing convergence as:
\begin{equation}
\kappa(\boldsymbol{x}) = \frac{\Lambda}{\Sigma_\textrm{cr}}\left(1+9.1\times10^{-2}\frac{\xi_0^2\boldsymbol{x}^2}{R_c^2}\right)^{-15/2}
\end{equation}
and choosing $x_c \equiv  R_c/\xi_0$ and $\boldsymbol{u} \equiv  \boldsymbol{x}/x_c$, our lensing potential takes the form
\begin{equation}
\label{eq-psi-x-SC}
    \psi(\boldsymbol{x}) = \frac{2\Lambda x_c^2}{\Sigma_\textrm{cr}}f(\boldsymbol{u})
\end{equation}

with 
\begin{equation}
\label{f_u}
    f(\boldsymbol{u}) = \frac{1}{\pi}\int d^2\boldsymbol{u}'\:\kappa(\boldsymbol{u}')\:\log|\boldsymbol{u} - \boldsymbol{u}'|.
\end{equation}

Following our choice of normalization, $\psi_{0}=1$, the quantity $\xi^\mathrm{SC}_0$ becomes:
\begin{equation}
    {\xi^\mathrm{SC}_0} = R_c \sqrt{\frac{2\Lambda}{\Sigma_\textrm{cr}}}
\end{equation}

\section{Rescaling of Solitonic Core}
\label{rescaling_SC}
{In this section, we outline the procedure for rescaling the solitonic core (SC) so that it can be embedded consistently within an NFW halo, forming a composite structure representative of an FDM halo. This rescaling allows us to compute the combined lensing potential using a unified normalization scale.
Following the approach in \cite{Brando_2025}, we adopt the NFW characteristic scale as $\xi^\mathrm{NFW}_0 = \sqrt{4 G d_\mathrm{eff} M_{Lz}}$ .
Starting from Eq.~\ref{eq-psi-x-SC}, the lensing potential of the SC can be expressed as
\begin{equation}
\label{eq-psi-x-SC-rescaling_1}
     \psi(\boldsymbol{x}) = \left(\frac{2\Lambda}{\Sigma_\textrm{cr}}f(\boldsymbol{u}) \right)\left(\frac{ R_c}{\xi^{\rm{SC}}_0}\right)^2 \times \left(\frac{\xi^\mathrm{NFW}_0}{\xi^\mathrm{NFW}_0}\right)^2.
\end{equation}
Here we define $x^{'}_c \equiv  R_c/\xi^{\rm{NFW}}_0$, which is the new rescaled length, and we rewrite the above equation (Eq.~\ref{eq-psi-x-SC-rescaling_1}) as 
\begin{equation}
\label{eq-psi-x-SC-rescaling_2}
     \psi(\boldsymbol{x}) = \frac{2\Lambda x^{'2}_c}{\Sigma_\textrm{cr}}f(\boldsymbol{u}) \times \left(\frac{\xi^\mathrm{NFW}_0}{\xi^\mathrm{SC}_0}\right)^2.
\end{equation}
where $f(\boldsymbol{u})$ is same as defined in Eq.~\ref{f_u} and the term $(\xi^\mathrm{NFW}_0 / \xi^\mathrm{SC}_0)^2$ is the one used to rescale the lensing potential $\psi(\boldsymbol{x})$.}

\nocite{*}

\bibliographystyle{JHEP}
\bibliography{biblio.bib}

\end{document}